\documentclass{article}
\usepackage{frascatiphys,here,graphicx,subfigure}
\begin{document}
\title{ 
STUDY OF STRUCTURE OF THE MASS GAP BETWEEN TWO SPIN MULTIPLETS
}
\author{
Takayuki Matsuki \\
{\em Tokyo Kasei University,
1-18-1 Kaga, Itabashi, Tokyo 173-8602, JAPAN} \\
Toshiyuki Morii \\
{\em Kobe University, 
Nada, Kobe 657-8501, JAPAN} \\
Kazutaka Sudoh \\
{\em
High Energy Accel. Res. Org.,
Tsukuba, Ibaraki 305-0801, JAPAN} \\
}
\maketitle
\baselineskip=11.6pt
\begin{abstract}
Studying our semirelativistic potential model and the numerical results, which
succeeds in predicting and reproducing recently discovered higher resonances of $D$,
$D_s$, $B$, and $B_s$, we find a simple expression for the mass gap between two spin
multiplets of heavy-light mesons, $(0^-,1^-)$ and $(0^+,1^+)$.
The mass gap
between chiral partners defined
by $\Delta M=M(0^+)-M(0^-)$ and/or $M(1^+)-M(1^-)$ is given by
$\Delta M=M(0^+)-M(0^-)=M(1^+)-M(1^-)\approx \Lambda_{\rm Q}-m_q$ in the limit of heavy
quark symmetry. We also study the case including $1/m_Q$ corrections.
\end{abstract}
\baselineskip=14pt
\section{Introduction}
\label{intro}
The discovery of the narrow $D_{sJ}$ particles by BaBar \cite{BaBar03} and CLEO
\cite{CLEO03} and soon confirmed by Belle \cite{Belle03} immediately reminded
people an effective theory approach proposed by Nowak et al. and others
\cite{Nowak93,Bardeen94,Ebert95,Deandrea98}.  
From this effective theory, they derived the Goldberger-Treiman relation for the mass
gap between chiral partners $0^{+}(1^{+})$ and $0^{-}(1^{-})$ 
instead of the heavy meson mass itself and predicted the mass gap
to be around $\Delta M=g_{\pi}f_{\pi}\approx 349$ MeV,
where $g_{\pi}$ is the coupling constant for $0^{+}\rightarrow 0^{-}+\pi$ and $f_{\pi}$
is the pion decay constant.

Since this mass gap between chiral partners in the case of $D_s$ agrees well
with the experiments (around 350 MeV), people thought that underling physics may
be explained by their $SU(3)$ effective Lagrangian \cite{Bardeen03,Harada04}.
However, when $(0^+,1^+)$ for $D$ meson were found by Belle and FOCUS, and later
reanalyzed by CLEO, their explanation needs to be modified.
Furthermore, what they originally predicted
could not be identified as any of heavy meson multiplets for $D$, $D_s$, $B$, and $B_s$.
In other words, the forumula can be applied equally for any of these heavy meson multiplets. 
Thus, it is required to find the mass gap formula, if it exists, which agrees well with
the experiments and explains the physical ground of its formula.

In this paper, using our semirelativistic potential model, we first give our formula for
the mass gap between chiral partners $0^{+}(1^{+})$ and $0^{-}(1^{-})$ 
for {\it any} heavy meson, $D$, $D_s$, $B$, and $B_s$,
among which the known mass gaps, i.e., the ones for $D$ and $D_s$, agree well with the experiments
although there is some ambiguities for $D$ meson data. Next we show how this mass gap depends
on a light quark mass $m_q$ for $q= u, d$, and $s$, where we neglect the difference between
$u$ and $d$ quarks.
Our formula naturally explain that the mass gap for $D$ is larger than that for $D_s$ and
predict the mass gaps for $B$ and $B_s$.
\section{Semirelativistic Quark Potential Model and Structure of Mass Gap}
\begin{figure}[t]
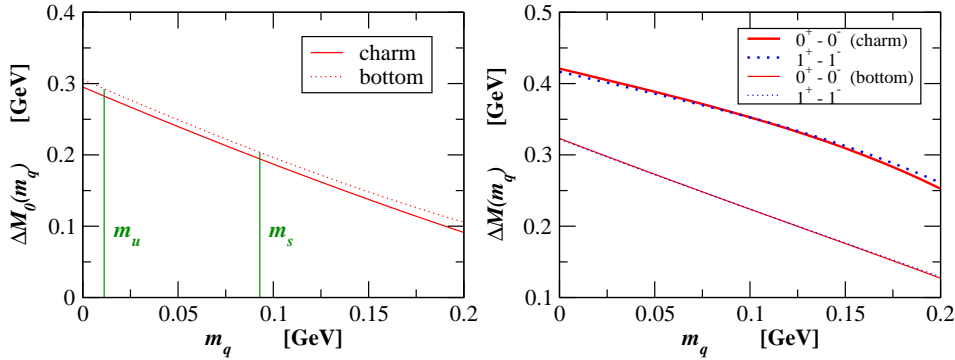

\includegraphics[scale=0.5,clip]{gap_0th.eps}
\includegraphics[scale=0.5,clip]{gap_1st1.eps}
\caption{Plots of the mass gap between two spin multiplets.
Light quark mass dependence is given. The horizontal axis is light quark mass $m_q$
and the vertical axis is the mass gap. Both the heavy quark limit $\Delta M_0$ (Left)
and the one with $1/m_Q$ corrections $\Delta M$(Right) are given.}
\label{fig-DeltaM}
\end{figure}

Mass for the heavy meson $X$ with the spin and parity, $j^P$, is expressed in our formulation
as \cite{Matsuki97}
\begin{eqnarray}
  M_X(j^P) = m_Q + E_0^k(m_q) + O\left(1/m_Q\right), \label{HMass}
\end{eqnarray}
where the quantum number $k$ is related to the total angular momentum $j$ and the
parity $P$ for a heavy meson as
\begin{eqnarray}
  j= |k|-1 {\rm ~~or~~} |k|, \quad P=\frac{k}{|k|}(-1)^{|k|+1}, \quad
  E_0^k(m_q)=E_0(j^P, m_q).
\end{eqnarray}
To begin with, we study the heavy meson mass without $1/m_Q$ corrections so that we
can see the essence of the mass gap.
States with the same $|k|$ value are degenerate in a pure chiral limit and without 
confining scalar potential,
which is defined as $m_q\rightarrow 0$ and $S(r) \rightarrow 0$ \cite{Matsuki05}. We consider
the scenario that a chiral symmetry breaking and a confinement take place in two steps. First
the degeneracy is broken due to gluon fields when $S(r)$ is turned on and confines quarks
into heavy mesons but keeping vanishing light quark mass intact.
In fact, in this limit our model gives
the mass gap between two spin multiplets $\Delta M \approx 300$ MeV as follows;
\begin{eqnarray}
  \Delta M &=& E_0(1^+,0) - E_0(1^-,0)=E_0(0^+,0) - E_0(0^-,0) \nonumber \\
  &=& 295.1 {\rm ~MeV~~for~} D, {\rm ~and~} D_s, \nonumber \\
  &=& 309.2 {\rm ~MeV~~for~} B, {\rm ~and~} B_s, 
\end{eqnarray}
This gap is mainly due to gluon fields which confines quarks into heavy mesons.
It is interesting that obtained values are close to
$\Lambda_{\rm QCD}\approx 300~{\rm MeV}$.
Next, turning on a light quark mass
which explicitly breaks a chiral symmetry, we have $SU(3)$ flavor
breaking pattern of the mass levels, i.e., mass of $D$ becomes different from that of
$D_s$ with the same value of $j^P$. Since we assume $m_u=m_d$, there still remains
$SU(2)$ iso-spin symmetry. Note that even after chiral symmetry is broken, there is still
degeneracy between members of a spin multiplet due to the heavy quark symmetry, i.e.,
$SU(2)_f\times SU(2)_{\rm spin}$ symmetry, with $SU(2)_f$ rotational flavor symmetry
and $SU(2)_{\rm spin}$ rotational spin symmetry.
By using the optimal values of parameters in Ref. \cite{Matsuki07}, which is listed
in Table \ref{parameter}, degenerate masses without $1/m_Q$ corrections 
for $D,~D_s$ and $B,~B_s$ mesons are calculated and presented in Table \ref{DegMassgap}.
Furthermore, by changing $m_q$ from 0 to 0.2 GeV, we have calculated the $m_q$
dependence of $\Delta M_0$ and have obtained Fig. \ref{fig-DeltaM}, in which $\Delta M_0$
is linearly decreasing with $m_q$. From Fig. \ref{fig-DeltaM}, we find that the mass gap between 
two spin multiplets for a heavy meson $X$ can be written as
\begin{eqnarray}
  \Delta M_0 &=& M_X(0^+) - M_X(0^-) = M_X(1^+) - M_X(1^-)
  =g_0\Lambda_{\rm Q}-g_1 m_q, \label{DM0} \\
  \Lambda _Q  &=& 300\;{\rm{MeV}},\;
  \left\{ {\begin{array}{*{20}c}
   {g_0  = 0.9836,\;g_1  = 1.080, \quad{\rm for~}D/D_s}  \\
   {g_0  = 1.017,\;g_1  = 1.089, \quad{\rm for~}B/B_s}  \\
  \end{array}} \right., \label{MassGap0}
\end{eqnarray}
where the values of $g_0$, and $g_1$ are estimated
by fitting the optimal line with Fig. \ref{fig-DeltaM}.
Since both $g_0$ and $g_1$ are very close to 1, we conclude that the mass
gap is essentially given by 
\begin{equation}
\Delta M_0=\Lambda_Q-m_q
\label{MassGap}
\end{equation}
Though the physical ground of this result is out of scope at present, Eq. (\ref{MassGap})
is serious, since it is very different from the one of an effective theory approach 
which gives the relation,
\begin{equation}
  \Delta M_0 = g_\pi\left(\left<\sigma\right>+m_q\right) .
\end{equation}
where $g_\pi$ is the Yukawa coupling constant between the heavy meson and a chiral
multiplet and is taken to be $g_\pi=3.73$ in \cite{Bardeen03}, and $\left<\sigma\right>=f_\pi$.
This expression is obtained
in the heavy quark symmetric limit and should be compared with our Eq.~(\ref{MassGap}). Instead
of minus sign for the term $m_q$ that we obtained, the authors of \cite{Bardeen94}
obtained plus sign as shown in the above equation. The same result is obtained even
if we use the nonlinear $\Sigma$ model \cite{Bardeen03}.
The result given by Eq.~(\ref{MassGap}) is exact when ${\cal O}\left(1/m_Q\right)$ terms are
neglected. As we will see later,  since $1/m_Q$ corrections are nearly equal
to each other for two spin doublets, the above equation (\ref{MassGap}) between two
spin multiplets holds approximately even with $1/m_Q$ corrections.
%
\begin{table}[t!]
\caption{Optimal values of parameters.}
\label{parameter}
\begin{tabular}{lcccccccc}
\hline
\hline
Params.
& ~~$\alpha_s^c$ & ~~$\alpha_s^b$ & ~~$a$ (GeV$^{-1}$) & ~~$b$ (GeV) \\
& ~~0.261$\pm$0.001 & ~~0.393$\pm$0.003 & ~~1.939$\pm$0.002 & ~~0.0749$\pm$0.0020 \\
& ~~$m_{u, d}$ (GeV) & ~~$m_s$ (GeV) & ~~$m_c$ (GeV) & ~~$m_b$ (GeV) \\
& ~~0.0112$\pm$0.0019 & ~~0.0929$\pm$0.0021 & ~~1.032$\pm$0.005 & ~~4.639$\pm$0.005 \\
\hline
& \# of data & \# of parameter & total $\chi^2$/d.o.f & \\
& 18 & 8 & 107.55 & \\
\hline
\hline
\end{tabular}
\end{table}

The reason why the mass gap can be written like Eq.~(\ref{MassGap0}) or (\ref{MassGap}) is
explained in our formulation. (See the details in Refs.\cite{Matsuki071} and \cite{Matsuki97}.)
\begin{table}[t!]
\caption{Degenerate masses of model calculations and their mass gap between $0^+(1^+)$
and $0^-(1^-)$ for $n=1$.}
\label{DegMassgap}
\begin{tabular}{lcccc}
\hline
\hline
~~~
& ~~$M_0(D)$ & ~~$M_0(D_s)$ & ~$M_0(B)$ & ~$M_0(B_s)$ \\
\hline
$0^-/1^-$
& ~~1784 & ~~1900 & ~5277 & ~5394 \\
$0^+/1^+$
& ~~2067 & ~~2095 & ~5570 & ~5598 \\
\hline
$0^+(1^+)-0^-(1^-)$
& ~~283 & ~~195 & ~293 & ~204 \\
\hline
\hline
\end{tabular}
\end{table}
%

\section{$1/m_Q$ Corrections}
\label{correction}
Next let us study the case when $1/m_Q$ corrections to the mass gap are taken into account.
Part of the results is given in \cite{Matsuki06}. In Table \ref{massgap}, we give our
numerical results in the cases of $n=1$ and $n=2$ (radial excitations). Values in brackets
are taken from the experiments. Our values seem to agree with the experimental ones
though the fit is not as good as the case for the absolute values of heavy meson masses.
We assume the form of the mass gap with the $1/m_Q$ corrections as follows.
\begin{equation}
  \Delta M = \Delta M_0 + \frac{c+d\cdot m_q}{m_Q}. \label{DM}
\end{equation}
Using Eq.(\ref{DM0}) for $D$ and $D_s$ mesons, i.e.
$\Delta M_0=g_0\Lambda_{\rm Q}-g_1m_q=295.1-1.080 m_q$, we obtain the values of the
parameters $c$ and $d$ for $D/D_s$ mesons given in Table
\ref{massgap}, which are given by
\begin{equation}
  c = 1.28\times 10^5~{\rm MeV^2}, \quad d = 4.26 \times 10^2~{\rm MeV}. 
  \label{cdParam}
\end{equation}
The term $c/m_Q$ lifts the constant $g_0\Lambda_Q$ about 100 MeV and the term
$d/m_Q$ gives deviation from -1 to the coefficient for $m_q$ in the case of $D/D_s$.

Applying this formula, Eq.~(\ref{DM}), to the case for $B/B_s$ with $m_Q=m_b$,
we obtain the mass gap as follows.
\begin{eqnarray}
  B(0^+)-B(0^-)&\approx& B(1^+)-B(1^-)\approx 322, \nonumber \\
  B_s(0^+)-B_s(0^-)&\approx & B_s(1^+)-B_s(1^-) 
  \approx 240 \quad {\rm MeV},
\end{eqnarray}
which should be compared with our model calculations, 321 and 241 MeV, in Table
\ref{massgap}. Thus the linear dependence of the mass gap on $m_q$ is also supported
in the case where the $1/m_Q$ corrections are taken into account.  The calculated
$m_q$ dependence  of $\Delta M$ with $1/m_Q$ corrections is presented in Fig. 2,
for $0<m_q<0.2$GeV.
\begin{table}[t!]
\caption{Model calculations of the mass gap. Values in brackets are taken from the
experiments. Units are MeV.}
\label{massgap}
\begin{tabular}{lcccc}
\hline
\hline
Mass gap ($n=1$)
& ~~$\Delta M(D)$ & ~~$\Delta M(D_s)$ & ~$\Delta M(B)$ & ~$\Delta M(B_s)$ \\
\hline
$0^+-0^-$
& ~~414 (441) & ~~358 (348) & ~322 & ~239 \\
$1^+-1^-$
& ~~410 (419) & ~~357 (348) & ~320 & ~242 \\
\hline
\hline
\end{tabular}
\end{table}
\begin{table}[t!]
\begin{tabular}{lcccc}
\hline
\hline
($n=2$)
& ~~$\Delta M(D)$ & ~~$\Delta M(D_s)$ & ~$\Delta M(B)$ & ~$\Delta M(B_s)$ \\
\hline
$0^+-0^-$
& ~~308 & ~~274 & ~~206 & ~~160 \\
$1^+-1^-$
& ~~350 & ~~327 & ~~216 & ~~171 \\
\hline
\hline
\end{tabular}
\end{table}
%


\section{Miscellaneous Phenomena}
{\it Global Flavor $SU(3)$ Recovery -- }
\label{revovery}
Looking at the mass levels of $0^+$ and $1^+$ states for the $D$ and $D_s$ mesons,
one finds that mass differences between $D$ and $D_s$ becomes smaller compared with those
of the $0^-$ and $1^-$ states. This can be seen from Table \ref{DDsmeson} and
was first discussed in Ref.\cite{Dmitrasinovic05} by Dmitra$\check{\rm s}$inovi\'c.
He claimed that considering
$D_{sJ}$ as a four-quark state, one can regard this phenomena as flavor $SU(3)$ recovery.
However, in our interpretation, this is not so as we have seen that this is caused by
the mass gap dependency on a light quark mass, $m_q$, as shown in Fig. \ref{fig-DeltaM}.
That is, when the mass of $D$ meson is elevated largely from the $0^-/1^-$ state to the
$0^+/1^+$ state, the mass of $D_s$ meson is elevated by 
about 100 MeV smaller than that of $0^-/1^-$
as one can see from Fig. \ref{fig-DeltaM}. In our interpretation, the $SU(3)$ is not
recovered since the light quark masses of $m_u=m_d$ and $m_s$ do not change their
magnitudes
when the transition from $0^-/1^-$ to $0^+/1^+$ occurs, and their values remain to be
$m_{u(d)}=11.2$ MeV and $m_s=92.9$ MeV, respectively, as presented in Table \ref{parameter}. \\
\begin{table*}[t!]
\caption{$D/D_s$ meson mass spectra for both the calculated and experimentally observed ones.
Units are MeV.}
\label{DDsmeson}
\begin{tabular}{@{\hspace{0.5cm}}c@{\hspace{0.5cm}}c
@{\hspace{1cm}}c@{\hspace{0.5cm}}|@{\hspace{0.5cm}}c@{\hspace{1cm}}c@{\hspace{0.5cm}}}
\hline
\hline
$^{2s+1}L_J (J^P)$ &
$M_{\rm calc}(D)$ & $M_{\rm obs}(D)$ &
$M_{\rm calc}(D_s)$ & $M_{\rm obs}(D_s)$ \\
\hline
\multicolumn{1}{@{\hspace{0.6cm}}l}{$^1S_0 (0^-)$} 
& 1869 & 1867
& 1967 & 1969 \\
\multicolumn{1}{@{\hspace{0.6cm}}l}{$^3S_1 (1^-)$} 
& 2011 & 2008
& 2110 & 2112 \\
\multicolumn{1}{@{\hspace{0.6cm}}l}{$^3P_0 (0^+)$} 
& 2283 & 2308
& 2325 & 2317 \\
\multicolumn{1}{@{\hspace{0.6cm}}l}{$"^3P_1" (1^+)$} 
& 2421 & 2427
& 2467 & 2460 \\
\hline
\hline
\end{tabular}
\end{table*}
\begin{table*}[t!]
\caption{$B/B_s$ meson mass spectra for both the calculated and experimentally observed ones.
Units are MeV.}
\label{BBsmeson}
\begin{tabular}{@{\hspace{0.5cm}}c@{\hspace{0.5cm}}c
@{\hspace{1cm}}c@{\hspace{0.5cm}}|@{\hspace{0.5cm}}c@{\hspace{1cm}}c@{\hspace{0.5cm}}}
\hline
\hline
$^{2s+1}L_J (J^P)$ &
$M_{\rm calc}(B)$ & $M_{\rm obs}(B)$ &
$M_{\rm calc}(B_s)$ & $M_{\rm obs}(B_s)$ \\
\hline
\multicolumn{1}{@{\hspace{0.6cm}}l}{$^1S_0 (0^-)$} 
& 5270 & 5279
& 5378 & 5369 \\
\multicolumn{1}{@{\hspace{0.6cm}}l}{$^3S_1 (1^-)$} 
& 5329 & 5325
& 5440 & $-$ \\
\multicolumn{1}{@{\hspace{0.6cm}}l}{$^3P_0 (0^+)$} 
& 5592 & $-$
& 5617 & $-$ \\
\multicolumn{1}{@{\hspace{0.6cm}}l}{$"^3P_1" (1^+)$} 
& 5649 & $-$
& 5682 & $-$ \\
\hline
\hline
\end{tabular}
\end{table*}
%
{\it Mass Gap of Heavy Baryons --}
\label{baryon}
When we apply our formula to the heavy-light baryons which include two heavy
quarks, $\left(ccs\right)$, $\left(ccu\right)$, $\left(bcs\right)$, $\left(bcu\right)$, $\left(bbs\right)$,
and $\left(bbu\right)$,
mass gaps between two pairs of baryons, like $\left(ccs\right)$ and $\left(ccu\right)$,
will be given by Eq.~(\ref{MassGap}) in the heavy quark symmetric limit and by
Eq.~(\ref{DM}) with $1/m_Q$ corrections where we have to replace $m_Q$ with
$m_{Q_1}+m_{Q_2}$. Here the isospin symmetry is respected since in our model
$m_u=m_d$.
This speculation is legitimized since
$QQ$ pair can be considered to be $3^*$ expression in the color $SU(3)$ space so that the
baryon like $QQq$ can be regarded as a heavy-light meson and our arguments expanded in this
paper can be applied \cite{Savage90,Ito93}.

\def\Journal#1#2#3#4{{#1} {\bf #2}, #3 (#4)}
\def\etal{{\it et al}.}
\def\NIM{Nucl. Instrum. Methods}
\def\NIMA{Nucl. Instrum. Methods A}
\def\NPB{Nucl. Phys. B}
\def\PLB{Phys. Lett. B}
\def\PRL{Phys. Rev. Lett.}
\def\PRD{Phys. Rev. D}
\def\PRO{Phys. Rev.}
\def\ZPC{Z. Phys. C}
\def\EPJ{Eur. Phys. J. C}
\def\EPJA{Eur. Phys. J. C}
\def\PR{Phys. Rept.}
\def\IJM{Int. J. Mod. Phys. A}
\def\PTP{Prog. Theor. Phys.}


\begin{thebibliography}{00}
\bibitem{BaBar03} BaBar Collaboration, B. Aubert \etal, 
  \Journal{\PRL}{90}{242001}{2003}.
\bibitem{CLEO03} CLEO Collaboration, D. Besson \etal, 
  \Journal{\PRD}{68}{032002}{2003}; 
\bibitem{Belle03} Belle Collaboration, P. Krokovny {\em et al.}, 
  \Journal{\PRL}{91}{262002}{2003}; Y. Mikami {\em et al.},
  \Journal{\PRL}{92}{012002}{2004}.
\bibitem{Nowak93} M. A. Nowak \etal, 
  \Journal{\PRD}{48}{4370}{1993}.
\bibitem{Bardeen94} W. A. Bardeen \etal,
  \Journal{\PRD}{49}{409}{1994}.
\bibitem{Ebert95} D. Ebert \etal,
  \Journal{\NPB}{434}{619}{1995};
  \Journal{\PLB}{388}{154}{1996}.
\bibitem{Deandrea98} A. Deandrea \etal, \Journal{\PRD}{58}{034004}{1998}.
\bibitem{Bardeen03} W. A. Bardeen \etal,
  \Journal{\PRD}{68}{054024}{2003}.
\bibitem{Harada04} M. Harada \etal,
  \Journal{\PRD}{70}{074002}{2004}.
\bibitem{Belle04} Belle Collaboration, K. Abe {\em et al.},
  \Journal{\PRD}{69}{112002}{2004}. 
\bibitem{Matsuki071} T. Matsuki \etal, hep-ph/0710.0325, to be published in Phys. Lett. B.
\bibitem{Matsuki97} T. Matsuki \etal, 
  \Journal{\PRD}{56}{5646}{1997}.
\bibitem{Matsuki05} T. Matsuki \etal,
  \Journal{\PLB}{606}{329}{2005}.
  hep-ph/0408326.
\bibitem{Matsuki07} T. Matsuki \etal,
  \Journal{\PTP}{117}{1077}{2007}.
\bibitem{Matsuki06} T. Matsuki \etal,
  \Journal{\EPJA}{31}{701}{2007}.
\bibitem{Dmitrasinovic05} V. Dmitra$\check{\rm s}$inovi\'c, 
  \Journal{\PRL}{94}{162002}{2005}.
\bibitem{Savage90} M. Savage  \etal, \Journal{\PLB}{248}{177}{1990}.
\bibitem{Ito93} T. Ito \etal, \Journal{\ZPC}{59}{57}{1993}.
%
\end{thebibliography}
\end{document}